\documentstyle[12pt]{article}
\newdimen\paperwidth
\newdimen\paperlength
\newdimen\margin
 \newdimen\vmargin
\textwidth=\paperwidth
\advance \textwidth by -2\margin
\advance\hoffset by -1truein
\advance\hoffset by \margin \textheight=\paperlength
\advance \textheight by -2\vmargin
\advance\voffset by -1truein
\advance\voffset by \vmargin
\advance\textheight by
-2\baselineskip
\begin{document}

\begin{titlepage} \title{ {\bf The Non-Linear Sigma Model }\\
{\bf and Spin Ladders} }

\vspace{2cm} \author{
{\bf Germ{\'a}n Sierra}\thanks{On leave of absence from the
Instituto de Matem{\'a}ticas y F{\'\i}sica Fundamental, C.S.I.C.,
Serrano 123, 28006-Madrid, Spain} \\ \mbox{}
\\
{\em
 Theoretische Physik - ETH-Honggerberg}\\ {\em
CH-8093, Switzerland } } \vspace{5cm}
\date{} \maketitle
\def\baselinestretch{1.3}

\begin{abstract}
The well known Haldane map from spin chains into the $O(3)$ non linear
sigma model is generalized to the case of spin ladders. This map
allows us to
explain the different qualitative behaviour between
even and odd ladders, exactly  in the same way
it explains the difference between integer and
half-integer spin chains. Namely, for even ladders the topological term
in the sigma model action is absent, while for odd ladders the
$\theta$ parameter, which multiplies the topological term,
is equal to $2 \pi S$, where $S$ is the spin of the ladder.
Hence  even ladders should have a dynamically generated
spin gap, while  odd ladders with half-integer spin
should stay gapless, and physically
equivalent to a perturbed $SU(2)_1$ Wess-Zumino -Witten model
in the infrared regime.
We also derive some consequences from the dependence of the sigma model
coupling constant on the ladder Heisenberg couplings constants.

\end{abstract}

\vspace{2cm} PACS numbers: 71.27.+a, 75.10 Jm, 05.50.+q.

\vskip3in
\end{titlepage}

\newpage
\def\baselinestretch{1.5} \noindent

\section{Introduction}

One of the most studied field theories in 2 dimensions are the
non-linear
sigma models. From a Particle Physics point of view
these models are ideal analogs of 4 dimensional quantum chromodynamics,
since they display  asymptotic freedom behaviour \cite{Polya},
dynamical mass generation, and existence of instantons \cite{Bela}.
In string theory the conformal invariant sigma models
are crucial to understand the on-shell properties on the strings
\cite{strings}.
In Solid State Physics the $O(3)$ sigma model plays also an important role in
understanding the properties of spin systems in various dimensions.
The map from spin chains into the sigma model
lead Haldane to the celebrated gap of antiferromagnetic
Heisenberg chains with integer values of the spin \cite{Hal},\cite{AffLH}.
The crucial parameter which controls the behaviour of the sigma model
is the angle $\theta$ that multiplies a topological
term into the action and which according to the map takes the
value $\theta= 2 \pi S$, where $S$ is the spin of the chain.
Haldane's prediction followed from the fact that the sigma models
with $\theta = 0 ({\rm mod} 2 \pi) $
are  massive field theories \cite{HKS},\cite{PW},\cite{ZZ1}.
This map was deduced
in the semiclassical limit where $S >> 1$,
but there is by now a clear experimental and theoretical evidence
of the existence of the gap \cite{Botet},\cite{Buyers}.
For half-integer values of the spin the prediction,
based on the gapless character of the spin 1/2 chain \cite{Bethe},
was that all these models should also be gapless. This has been confirmed by
numerical computations \cite{Zi}. The sigma model at $\theta = \pi$ has
also been proved to be massless \cite{Shan}.

Nowadays there is a better understanding of the sigma model at $\theta =\pi$
by means of the powerful techniques
of bosonization \cite{AffNA}, \cite{Schulz}, conformal field theory
and also from the factorized scattering
theory \cite{Fateev},\cite{ZZ2}. It has been
shown that the low energy physics of the $\theta = \pi$ model
is well described by the $SU(2)_1$
Wess Zumino Witten model \cite{Aff-Hal}, \cite{FOZ}.
The asymptotically free theory in the ultraviolet region,
which is described by two goldstone
modes, becomes in the infrared region
a  $SU(2)_1$ WZW model \cite{Fateev}.
This RG flow  satisfies the Zamolodchikov c-theorem \cite{Zc}.
An alternative description of the sigma model for low energies
is given by a marginal-irrelevant
perturbation of the WZW model by the product of the two chiral $SU(2)$
currents \cite{Aff-Hal}.
In this manner one can do perturbative calculations around
the conformal point \cite{Cardy}.
It is interesting to observe that
there are not relevant perturbations of the $SU(2)_1$ WZW model, 
which explains why this conformal field theory characterizes the universality
class of a large variety of spin systems.

We shall show in this paper that the sigma model
methods can be extended
to spin ladders, which will then allow us to consider certain questions
arising in this subject.
Spin ladders  are arrangements of $n_{\ell}$
parallel spin chains with nearest
neighbour Heisenberg couplings between the spins
along and across the chains. These spin systems are
interpolating structures between 1 and 2 dimensions. The interest on spin
ladders increased enormously when experimentalist discovered materials
like $(VO)_2 P_2 O_7$ \cite{VOP}
and  $Sr_{n-1} Cu_{n+1} O_{2n}$
\cite{SrCuO}, whose magnetic and electronic structure was
analysized in \cite{Rice}.
Hence from an experimental and
theoretical point of view the spin ladders have become
a place where to
test different ideas concerning strongly
correlated systems ( see reference \cite{DaRi} for a review on the subject).

A central question in the study of spin ladders is their different
qualitative behaviour as a function of the number of legs $n_{\ell}$.
The main conjecture is that ladders with even number of legs
have a finite spin gap and short distance correlations while
the odd ladders have gapless spin excitations,
and power-law correlations.
Many authors have contributed to clarify this
question, and despite of some initial
controversies, it  is clear by now that it must be
correct \cite{Dag}-\cite{Shelton}.
In this paper  we shall give further support
of this conjecture  using sigma model techniques,
which will clearly
show the
topological nature of the mechanism underlying
the existence or absence of a spin gap in the ladder's spectrum.
Our proof is
an extension of the Haldane's result from
the chain to the ladder. Indeed,
we shall show that the low lying modes of the ladder are described by a
sigma model with a value of the angle $\theta$ equal to zero for the even
chains and equal to $2 \pi S$ for the odd ones. Hence from this result and
the well known properties of the sigma model
at $\theta=0$ and $\pi$ we prove
the above conjecture. This kind of proof has been suggested in
\cite{Kh} on the basis of the
2d formulation  of the sigma model due to Haldane
, who showed  the absence of the
topological term in the 2d Heisenberg antiferromagnet
\cite{Hal2}.
However our approach to the problem is not really 2 dimensional since
we take into account the specific
nature of the ladders (i.e. objects  in between 1d and 2d).
This different treatment  is made clear by the nature of
the sigma model variables that we use
which are one dimensional fields.

In the case of odd ladders with half-integer spins there is an extension
of the Lieb-Mattis-Schultz theorem due to Affleck which states that in the
infinite length limit , either the ground state is degenerate or else there
are gapless excitations \cite{LMS}, \cite{LMSA}.
Our results, together with those of references
\cite{Rice},\cite{White},\cite{Gopalan},
\cite{Rei}, \cite{Hatano} confirm that
the last possibility is the one realized by the spin ladders, the spin chain
being the particular case  $n_{\ell}=1$.
The LMSA theorem works under very general circunstances, which is a
manifestation of the topological nature of odd ladders and half-integer spins.
To make this more transparent
we shall consider spin ladders with arbitrary values
of the interchain and intrachain coupling constants. Our results
concerning the nature of the gap
will be independent of the precise values taken by these parameters.
We want to mention here  another topological interpretation of the
difference between even and odd ladders given in \cite{White}
in the framework of the RVB picture \cite{Ander}.
According to \cite{White} the even ladders are short range RVB
systems which have a gap due to the confinement of topological deffects,
while the odd ladders are long range RVB systems with no confinement
and consequently no gap. It would be interesting to analize
the relation between these two topological interpretations.

\section{AF Spin ladders: goldstone modes}

The Hamiltonian of a spin ladder with $n_{\ell}$ legs of length N is given by,

\begin{eqnarray}
& H_{\rm ladder}= H_{\rm leg} + H_{\rm rung}&  \label{1}
\\   H_{\rm leg }= & \sum_{a=1}^{n_{\ell}} \sum_{n=1}^N  J_a \;
{\bf S}_a(n) \cdot {\bf S}_a(n+1) & \nonumber \\
H_{\rm rung }= &  \sum_{a=1}^{n_{\ell}-1} \sum_{n=1}^N  J'_{a,a+1} \;
{\bf S}_a(n) \cdot {\bf S}_{a+1}(n)\nonumber  &
\end{eqnarray}

\noindent
where ${\bf S}_a(n)$ are spin-S matrices
located in the $a^{th}$ leg
at the position $n=1, \dots,N$. We consider periodic boundary
conditions along the legs ( $ {\bf S}_a(n)=  {\bf S}_a(n+N) )$.
The only condition we shall impose on the coupling constants
$J_a$ and $J_{a,a+1}'$ is that they are positive, which guarantee
that  $H_{\rm ladder}$
posseses, in the classical limit ( $S \rightarrow \infty$),
a minima given by the antiferromagnetic vacuum solution,

\begin{equation}
{\bf S}_a^{\rm class}(n)  = (-1)^{a+n}\; S \; {\bf z}
\label{2}
\end{equation}

\noindent
where ${\bf z}$ is the unit vector in the vertical direction.
The solution (\ref{2}) breaks the $O(3)$ rotational invariance
of $H_{\rm ladder}$ down to the subgroup  $O(2)$ of rotations
around the ${\bf z}-$axis. Consequently there should appear two goldstone
modes associated to the broken generators $S^x$ and $S^y$ \cite{AffLH}.
In the thermodynamic limit
where $N \rightarrow \infty$, with $n_{\ell}$ keept fixed,
the spectrum of the
Hamiltonian (\ref{1}) becomes essentially one dimensional, despite of
its 2d origin, and hence one
expects that the quantum corrections will restore the $O(3)$ symmetry, as
it happens for the usual Heisenberg chain.
Our strategy will then  parallel  the one used for the study of
spin chains. We shall only consider the massless degrees
of freedom associated to the two goldstone modes and later on
we shall consider their interaction in the framework of the sigma model.
An  direct way to find the
goldstone modes, which are nothing but
spin waves, is through the linearized approximation of the equation
of motion of the spins \cite{AffLA}.

The evolution equation of the spin operators of the ladder are given by,

\begin{eqnarray}
& \frac{d {\bf S}_a(n)}{d t} = i \left[ H_{\rm ladder}, {\bf S}_a(n)
\right]= &
\label{3} \\
& - {\bf S}_a(n) \times \left[ J_a  \left( {\bf S}_a(n+1)+
{\bf S}_a(n-1) \right) + J'_{a,a+1}  {\bf S}_{a+1}(n) +
J'_{a,a-1}  {\bf S}_{a-1}(n)
\right] \nonumber &
\end{eqnarray}

\noindent
This equation is valid for any $a=1, \dots , n_{\ell}$ with the
convenium  $J'_{0,1}= J'_{n_{\ell},n_{\ell}+1}=0$ and
$J'_{a,b}= J'_{b,a}$.
Expanding  ${\bf S}_a(n) $
around the classical solution (\ref{2}),

\begin{equation}
{\bf S}_a(n) = {\bf S}_a^{\rm class}(n) + {\bf s}_a(n)
\label{4}
\end{equation}

\noindent
one gets in the linearized approximation,

\begin{eqnarray}
& \frac{d {\zeta}_a(n)}{d t} = i (-1)^{a+n+1} S
\left[ J_a \left(  {\zeta}_a(n+1)+  {\zeta}_a(n-1)
+2  {\zeta}_a(n) \right)
+\sum_b K_{a,b}^+ \; {\zeta}_b(n) \right]&
\label{5}
\end{eqnarray}

\noindent
where $ {\zeta}_a(n) = {s}_a^x(n) + i {s}_a^y(n)$ and
the matrix $K^+_{a,b}$ together with a matrix $K^-_{a,b}$, which
we shall use later on, are defined as follows,

\begin{equation}
K^{\pm}_{a,b} =  \left\{ \begin{array}{cl}
J'_{a,a+1} + J'_{a,a-1} & a=b  \\ & \\  \pm J'_{a,b} & |a-b|=1
\end{array} \right.
\label{6}
\end{equation}

The solution of eqs.(\ref{5}) are given by plane waves,

\begin{equation}
{\zeta}_a(n) = e^{i( \omega t  + k n)}
\left( \psi_a(k) + (-1)^{a+n+1} \phi_a(k) \right)
\label{7}
\end{equation}

Introducting (\ref{7}) into (\ref{5}) one gets,

\begin{eqnarray}
& \frac{\omega}{S} \psi_a(k) = 4 \sin^2 \frac{k}{2} \;
J_a \;  \phi_a(k) + \sum_b K^-_{a,b}\; \phi_b(k) & \label{8} \\
& \frac{\omega}{S} \phi_a(k) = 4 \cos^2 \frac{k}{2} \;
J_a \; \psi_a(k) + \sum_b K^+_{a,b} \; \psi_b(k) & \nonumber
\end{eqnarray}

These equations have massless and massive modes
in the limit $N \rightarrow \infty$ ( periodicity along the legs implies
$k= 2 \pi m/N$ ,$m =0,1, \cdots , N-1$).
These modes  can be obtained by expanding
$ \omega(k),\psi_a(k)$ and $\phi_a(k)$ in  powers of the momenta $k$.
For the massless modes this expansion reads,

\begin{eqnarray}
\omega = & { v} k + O(k^3) & \nonumber \\
\psi_a(k) =& k A_a + O(k^3) & \label{9}
\\
\phi_a(k) =& B_a + k^2 C_a + O(k^4) \nonumber
\end{eqnarray}

The equations for the coefficients ${ v}, A_a, B_a $ and $C_a$
follows from (\ref{8}),

\begin{eqnarray}
0 =& \sum_b K^-_{a,b} \; B_b & \label{10} \\
\frac{ { v} }{S} B_a =& \sum_b L_{a,b} \;  A_b & \label{11} \\
\frac{ { v} }{S} A_a = & J_a B_a +
\sum_b K^-_{a,b} \; C_b & \label{12}
\end{eqnarray}

\noindent
where we have introduced yet another matrix $L_{a,b}$ given by,

\begin{equation}
L_{a,b} = 4 J_a \; \delta_{a,b}  + K^+_{a,b}
\label{13}
\end{equation}

\noindent
which is going to play an important role in the construction.
Notice that
$L_{a.b}$ is a positive definite matrix.

The solution of equation (\ref{10}) is uniquely given by
$B_a= B \;\; \forall a$. This result follows from the observation
that $K^-_{a,b}$ is a generalized
incidence matrix associated to a graph consisting of $n_{\ell}$ points
labelled by $a$ and links joining the points $a$ and $b$ whenever
$J'_{a,b}$ is non null. Since the graph is connected (i.e.
$J'_{a,b} $ is a non singular matrix)  there
is a unique vector satisfying equation (\ref{10}).
Connectedness of the graph simply means that the ladder
cannot be split into two or more subladders.
This graph together with its incidence matrix contains
all the information of  the rungs of the spin ladder relevant to our problem.

The solution of (\ref{11}) is given by,

\begin{equation}
A_a = \frac{ { v}}{S} B \sum_b L^{-1}_{a,b} \label{14}
\end{equation}

\noindent
where we have inverted the matrix $L$ ( recall that $L$ is positive definite).
To solve equation (\ref{12}) we proceed  in two steps. First of all
we sum over
the index $a$ in (\ref{12}) and use the fact that
$\sum_a K^-_{a,b} =0 $, to get rid of the term proportional to
$C_a$.
This give us an equation for the spin wave velocity ${ v}$,
which with the help of (\ref{14}) can be written as,

\begin{equation}
\left( \frac{ { v} }{S} \right)^2 =
\frac{ \sum_a J_a }{ \sum_{a,b} L^{-1}_{a,b} }
\label{15}
\end{equation}

Both the numerator and the denominator of (\ref{15}) are positive
which yields a real value for ${ v}/S$.

The solution of (\ref{12})
for the vector $C_a$ constitute in fact a one parameter
family of solutions given by $C_a + x B_a$ with $x$ arbitrary.
This  freedom reflects the linearity of
equations (\ref{8}). Multiplying the whole
solution (\ref{7}) by a $k$ dependent factor produces a term
of the form $k^2 B_a$ in (\ref{9}).
The ``transverse'' components of $C_a$ can then be obtained by inverting the
matrix $K^-$ matrix in the subspace orthogonal to its zero eigenvector.

Next we shall briefly consider the massive modes. The value of the gap
$\Delta = \omega(k=0)$ can be simply obtained setting $k=0$ in
(\ref{8}),

\begin{eqnarray}
&\frac{ \Delta}{S} \psi_a(0) = \sum_b K^-_{a,b}  \phi_b(0) & \label{16} \\
&\frac{ \Delta}{S} \phi_a(0) = \sum_b L_{a,b}  \psi_b(0) & \nonumber
\end{eqnarray}

Hence combining both equations we get that $(\Delta/S)^2$ is given
by the eigenvalues of the matrix $L K^-$ or alternatively
$K^- L$. One of this eigenvalues
is zero and corresponds to the massless mode studied above and the others
are all non zero and positive corresponding to the massive modes.

\section{ $\sigma-$model mapping}

Let us recall how one maps the Heisenberg spin chain
into the 1d $\sigma-$model
(we shall follow closely reference \cite{AffLH}). The spin wave analysis
shows that the spin operators ${\bf S}(n)$ has two smooth components
centered and momenta $k=0$ and $k=\pi$, which can be identified with
the total angular momenta ${\bf l}$ and the staggered field ${\bf \varphi} $
respectively. The relation between these operators can be written as
follows,

\begin{eqnarray}
{\bf S}(2n)= & {\bf l}(x) - S {\bf \varphi}(x) & \label{17} \\
{\bf S}(2n+1)= & {\bf l}(x) + S {\bf \varphi}(x) & \nonumber
\end{eqnarray}

\noindent
where $x= 2n+ \frac{1}{2}$ is the midpoint coordinate of the block formed
by the points $2n$ and $2n+1$. Inverting eq.(\ref{17}) one gets,

\begin{eqnarray}
{\bf l}(x) = &  \left( {\bf S}(2n+1) + {\bf S}(2n) \right)/2
& \label{18} \\
{\bf \varphi}(x) = &  \left( {\bf S}(2n+1) - {\bf S}(2n)
\right)/2S
& \nonumber
\end{eqnarray}

${\bf l}$ and ${\bf \varphi}$  satisfy the following
commutation  relations,

\begin{eqnarray}
\left[ {\bf l}^i(x) , {\bf l}^j(y) \right] =& i \frac{ \delta_{x,y}}{2}
\; \epsilon^{ijk}\; {\bf l}^k(x) & \nonumber \\
\left[ {\bf l}^i(x) , {\bf \varphi }^j(y) \right] =&
i \frac{ \delta_{x,y}}{2}
\; \epsilon^{ijk}\; {\bf \varphi }^k(x) & \label{19} \\
\left[ {\bf \varphi }^i(x) , {\bf \varphi }^j(y)
\right] =& i  \delta_{x,y}
\; \epsilon^{ijk}\; {\bf l}^k(x)/2S^2 \rightarrow 0 & \nonumber
\end{eqnarray}

\noindent
which can be derived from the $SU(2)$ algebra satisfied by the
spin operators ${\bf S}(n)$.
The term $ \frac{ \delta_{x,y}}{2}$
is the lattice version of the Dirac's delta function
$\delta(x-y)$. The factor two in the denominator is simply the lattice
spacing of the 2-block arrangement of the chain.
The fact that ${\bf S}(n)$ are spin-S matrices
satisfying the relation ${\bf S}^2(n) = S(S+1)$, imply
two additional equations for ${\bf l}$ and ${\bf \varphi}$, namely,

\begin{eqnarray}
&{\bf \varphi}^2 =1 - {\bf l}^2/S^2 + O(1/S) \rightarrow 1 & \label{20} \\
& {\bf l} \cdot {\bf \varphi} = 0 & \nonumber
\end{eqnarray}

Equations (\ref{19}) and (\ref{20}) are the standard relations
satisfied by the $\sigma-$field ${\bf \varphi}$ and the angular
momenta ${\bf l}$. Introducing eqs.(\ref{17}) into the
spin chain Heisenberg Hamiltonian one gets, taking the continuum limit
and eliminating higher
derivatives terms, the standard $\sigma-$model Hamiltonian,

\begin{equation}
H_{\sigma } = \frac{ v_{\sigma} }{2}
\int dx  \left[ g \left( {\bf l}^2 - \frac{ \theta}{ 4 \pi}
{\bf \varphi}'  \right)^2  + \frac{1}{g} {\bf \varphi}'^2
\right]
\label{21}
\end{equation}

\noindent
where ${\bf \varphi}' = \partial_x {\bf \varphi}$.
The theta parameter, coupling constant
and spin velocity take the following values,

\begin{equation}
\theta = 2 \pi S ,\; \; g= \frac{2}{S}, \; \; v_{\sigma}= 2 J S
\;\;\;\;  (n_{\ell} =1).
\label{22}
\end{equation}

The Hamiltonian (\ref{21}) can be obtained from the 2d
$\sigma-$model Lagrangian,

\begin{equation}
L = \frac{1}{ 2 g} \partial_{\mu} {\bf  \varphi}
\cdot \partial_{\mu} {\bf  \varphi} + \frac{ \theta}{ 8 \pi} \;
\epsilon^{\mu \nu} \; {\bf \varphi} \cdot \left(
\partial_{\mu} {\bf  \varphi} \times  \partial_{\nu}
{\bf  \varphi} \right)
\label{23}
\end{equation}

It is our aim to generalize
the previous construction to spin ladders. First of all we shall
divide the ladder into blocks of two consecutive rungs and define
$\sigma-$model variables for each of them. The spin wave analysis
of the previous section suggest the following ansatz,

\begin{eqnarray}
{\bf S}_a(2n) = & A_a {\bf l} + {\bf l}_a +
S (-1)^{a}  ( {\bf \varphi} + {\bf \varphi}_a ) & \label{24} \\
{\bf S}_a(2n+1) = & A_a {\bf l} + {\bf l}_a +
S (-1)^{a+1}  ( {\bf \varphi} + {\bf \varphi}_a )\nonumber &
\end{eqnarray}

\noindent
where ${\bf l}$ and ${\bf \varphi}$ are the candidates for the
$\sigma $ variables and  ${\bf l}_a$ and ${\bf \varphi}_a$ are some extra
slowly varying fields needed to
match the number of degrees of freedom in both sides
of (\ref{24}). For these to be the case we shall impose the following
``transversality conditions'' on  ${\bf l}_a$ and ${\bf \varphi}_a$,

\begin{eqnarray}
& \sum_a {\bf l}_a = 0 & \label{25} \\
& \sum_a {\bf \varphi}_a=0 & \nonumber
\end{eqnarray}

\noindent
which are only needed for $n_{\ell} > 1$.
Using (\ref{25}) we can express  ${\bf l}$ and ${\bf \varphi}$
in terms of the spin operators as follows,

\begin{eqnarray}
{\bf l}(x) = &   \sum_a \left[
{\bf S}_a(2n + 1) + {\bf S}_a(2 n)
\right]/\left(  2 \sum_b A_b \right) &  \label{27} \\
{\bf \varphi}(x) = &   \sum_a
(-1)^{a+1}
\left[
{\bf S}_a(2n + 1) - {\bf S}_a(2 n)
\right]/\left( 2S n_{\ell} \right) &  \nonumber
\end{eqnarray}

We want ${\bf l}$ and ${\bf \varphi}$ to satisfy the algebraic relations
(\ref{19}), which can be achieved imposing,

\begin{equation}
\sum_a A_a =1  \Rightarrow A_a =
\frac{ \sum_b L_{a,b}^{-1} }{ \sum_{c,d} L^{-1}_{c,d} }
\label{28}
\end{equation}

\noindent
where we have used eq.(\ref{14}).
Similarly ${\bf l}$ and ${\bf \varphi}$ as given by (\ref{27})
satisfy eqs. similar to (\ref{20}),

\begin{eqnarray}
&{\bf \varphi}^2 =1 + O(1/S n_{\ell}) & \label{29} \\
& {\bf l} \cdot {\bf \varphi} = O(1/S n_{\ell}) & \nonumber
\end{eqnarray}

Hence in the limit $S n_{\ell}>>1  $ we obtain the constraints
which define the $\sigma-$model. From (\ref{29})
it seems that the expansion parameter that we are employing
is $S n_{\ell}$ rather than $S$. If this is correct then considering
higher spins is equivalent, from the sigma model point of view,
to considering ladders with many legs. We shall return
later on
to this
suggestion.
Another point which  is worth to mention is that
${\bf l}(x)$ and ${\bf \varphi}(x)$ represent total
angular momenta
and staggered magnetization of the rung taken as a whole.
This is why they are 1d densities depending
only on the single coordinate
$x$.
Given the relations (\ref{24}) we can now write the spin
ladder hamiltonian (\ref{1}) in the variables
${\bf l},{\bf \varphi}, {\bf l}_a$ and ${\bf \varphi}_a$
as follows,

\begin{eqnarray}
&H_{ladder} = \int \frac{dx}{2}  \left\{
\sum_{a,b} L_{a,b}  \left( A_a A_b \; {\bf l}^2 + {\bf l}_a \; {\bf l}_b
\right)  \right. & \nonumber \\ & & \nonumber \\
&+ 2 S^2 \sum_a J_a \left( {\bf \varphi}' + {\bf \varphi}'_a \right)^2
+ \sum_{a,b} K^-_{a,b}\; {\bf \varphi}_a \; {\bf \varphi}_b  &
\label{30} \\ & & \nonumber \\
& \left. +2 S \sum_a (-1)^{a} J_a \left[ \left( A_a {\bf l} + {\bf l}_a \right)
\left(  {\bf \varphi}' + {\bf \varphi}'_a \right)
+ \left( {\bf \varphi}' + {\bf \varphi}'_a \right)
\left( A_a {\bf l} + {\bf l}_a \right) \right] \; \right\} & \nonumber
\end{eqnarray}

To derive (\ref{30}) we have used (\ref{11}), (\ref{25}) and
the following formula,

\begin{eqnarray}
& \left( A_a {\bf l} + {\bf l}_a \right)^2 +
S^2  \left( {\bf \varphi} + {\bf \varphi}_a \right)^2 = S(S+1)
& \label{31} \\
&\left( A_a {\bf l} + {\bf l}_a \right)
\left( {\bf \varphi} + {\bf \varphi}_a \right)=0 &
\nonumber
\end{eqnarray}

\noindent
which is a consequence of the relations
${\bf S}_a^2(n) = S( S+1)$ (since we are working in the semiclassical
limit $S>>1$ we shall keep only the highest power in S).
To decouple the fields $\varphi$ and $\varphi_a$ 
in (\ref{30}) we have to choose the same value of
$J_a$ for all the legs. Indeed upon this condition the cross
term $ \sum_a J_a \varphi` \varphi`_a$  
in (\ref{30}) vanishes as a consequence of (\ref{25}). We thus obtain
that $\varphi$ is a massless field while the fields $\varphi_a$ are
massive. 
Let us now concentrate on the massless field,
whose  Hamiltonian reads,

\begin{eqnarray}
&H_{\rm ladder}^{(\rm massless)} = \int \frac{dx}{2}
\left[ \left( \sum_{a,b} L_{a,b}  A_a A_b \right)
{\bf l}^2 + 2 S^2 \sum_a J_a {\bf \varphi}^2 + 2S \sum_a (-1)^a J_a A_a
\left( {\bf l}\; {\bf \varphi}' + {\bf \varphi}'\; {\bf l}
\right) \right] &
\label{32}
\end{eqnarray}

This is precisely the $\sigma-$model Hamiltonian given in (\ref{21})
with an appropiate identification of $\theta, g $ and $ v$.
Let us first consider $\theta$ whose values is given by,

\begin{equation}
\theta = 8 \pi S \frac{ \sum_a (-1)^{a+1} J_a A_a }{
\sum_{b,c} L_{b,c} A_b  A_c }
\label{33}
\end{equation}

A simplification of (\ref{33}) is achieved using (\ref{28}),

\begin{equation}
\theta = 8 \pi S \sum_{a,b} (-1)^{a+1} J_a L_{a,b}^{-1}
\label{34}
\end{equation}

A convenient way of writing (\ref{34}) is by means of two
$n_{\ell}$-dimensional
vectors $|F>$ and $|AF>$  defined as follows,

\begin{equation}
|F> = \frac{1}{\sqrt{ n_{\ell} }}
\left( \begin{array}{c}  1 \\ 1 \\ \vdots \\ 1 \end{array}
\right), \;\;
|AF> = \frac{1}{\sqrt{ n_{\ell} }}
\left( \begin{array}{c}  1 \\ -1 \\ \vdots \\
(-1)^{n_{\ell}+1}  \end{array}
\right)
\label{35}
\end{equation}

\noindent
whose scalar product is,

\begin{equation}
<AF|F> = \left\{ \begin{array}{cl}
0 & n_{\ell}: {\rm even} \\ & \\
{1}/{n_{\ell}} & n_{\ell}:{\rm odd}
\end{array} \right.
\label{36}
\end{equation}

Using (\ref{35}) we write (\ref{34}) in matrix notation as,

\begin{equation}
\theta = 8 \pi S n_{\ell} < AF|\; {\bf J}\; {\bf L}^{-1}\; |F>
\label{37}
\end{equation}

\noindent
where ${\bf J}$ is a diagonal matrix whose  entries are $J_a$.
Recalling the well known operator identity,

\begin{equation}
\frac{1}{A+B} =  \frac{1}{A} - \frac{1}{A}\; B \; \frac{1}{A+B}
\label{38}
\end{equation}

\noindent
we transform (\ref{37}) into,

\begin{equation}
\theta = 2 \pi S n_{\ell}
< AF| \left( {\bf 1}- {\bf K}^+  \frac{ 1}{
4 {\bf J} + {\bf K}^+ } \right) |F>
\label{39}
\end{equation}

Then noticing that the vector $|AF>$ is annihilated by ${\bf K}^+$
and using eq.(\ref{36}) we arrive finally at,

\begin{equation}
\theta= \left\{ \begin{array}{cl}
0 & n_{\ell}:{\rm even} \\ & \\
2 \pi S & n_{\ell}:{\rm odd}
\end{array} \right.
\label{40}
\end{equation}

Taking into account that $\theta$ is defined modulo $2 \pi$ then
we can write eq.(\ref{40}) simply as,

\begin{equation}
\theta = 2 \pi S n_{\ell}
\label{40b}
\end{equation}

This result is valid for any value of the coupling constants
$J_a$ and $J_{a,a+1}$ as long as they are non vanishing.  This confirms
the topological nature of the result in total agreement with the LMSA
theorem. Moreover the derivation of (\ref{40b}) suggest that the $\theta$
term for spin ladders is related to the transition amplitude
from ferromagnetic to antiferromagnetic configurations along the rungs
(\ref{36}). A path integral derivation of (\ref{40b}), along
the lines of \cite{Hal2} , would probably throw
some light on this interpretation.

Next we shall give the expressions of the $\sigma-$model coupling
constant $g$ and the velocity $v_{\sigma}$ in terms of the ladder
parameters,

\begin{equation}
g^{-1} = S \left[ 2 \sum_{a,b,c} J_a \; L^{-1}_{b,c}\;  - \frac{1}{4}
\delta_{n_{\ell}} \right]^{1/2}
\label{41}
\end{equation}

\begin{equation}
\left( \frac{v_{\sigma} }{ S} \right)^2 =
2 \frac{ \sum_a J_a}{ \sum_{b,c} L^{-1}_{b,c}}
- \delta_{ n_{\ell} } \frac{ 1}{ \left( 2 \sum_{b,c} L^{-1}_{b,c}
\right)^2 }
\label{42}
\end{equation}

\noindent
where $\delta_{n_{\ell}}$ is equal to 1 (or 0) whenever $n_{\ell}$
is odd (or even).

Compairing eqs.(\ref{42}) and (\ref{15})
for $n_{\ell}=1$ we get that $v_{\sigma} =v$, but for $n_{\ell} >1$
the two velocities, $v_{\sigma}$ and
$v$, do not coincide (for $n_{\ell}$ even one has
$v_{\sigma} = \sqrt{2}\; v$).
We interpret this fact as a
kind of interference effect between the legs of the ladder which makes
the effective spin velocity $v_{\sigma }$ to differ from the spin wave
velocity $v$.
The most interesting case
for practical applications is when
$J_a = J$ and $J'_{a,a+1} = J'$
$\forall a$.
We shall give below the values of $g$
and $v_{\sigma}$ in this situation.

\begin{equation}
g^{-1} = S \left[ \frac{n_{\ell}^2 }{ 2} f(n_{\ell}, J'/J)
- \frac{1}{4}\; \delta_{n_{\ell}} \right]^{1/2}
\label{43}
\end{equation}

\begin{equation}
 v_{\sigma}  = \frac{4J}{ n_{\ell} \;
g \; f(n_{\ell}, J'/J) }
\label{44}
\end{equation}

The function $f(n_{\ell}, J'/J) $ appearing in these formulae is defined
as,

\begin{eqnarray}
& f(n_{\ell},J'/J) = < F| \; \left( 1 + \frac{ K^+ }{4J}
\right)^{-1} \;
|F> = & \nonumber \\ & & \label{45}  \\
& \frac{1}{ n_{\ell}^2 } \left[
\delta_{n_{\ell} }
+ 2 \; \sum_{ m=1,3, \cdots, n_{\ell}-1}
\left( \sin \frac{\pi m }{ 2 n_{\ell}} \right)^{-2}
\left( 1 + \frac{J'}{J} \;
\cos^2 \frac{ \pi m}{ 2 n_{\ell}} \right)^{-1} \right] &
\nonumber
\end{eqnarray}

In the particular cases where $n_{\ell}= 2$ and 3 we obtain,

\begin{equation}
{\rm for} \;\; n_{\ell} =2 \;\;\; \left\{ \begin{array}{l}
g= \frac{1}{S \sqrt{2}} \; \left( 1 + \frac{ J'}{2J} \right)^{1/2} \\ \\
v_{\sigma}= 2 \sqrt{2} S J \; \left( 1 + \frac{ J'}{2J} \right)^{1/2}
\end{array} \right.
\label{46}
\end{equation}

\begin{equation}
{\rm for}\;\; n_{\ell} =3 \;\;\; \left\{ \begin{array}{l}
g= \frac{2}{S } \;
\left( \frac{  1 + \frac{3 J'}{4 J}}{ 17 + \frac{3J'}{4J}} \right)^{1/2}\\  \\
v_{\sigma}= \frac{ 2  J  S}{3} \; \frac{ \left[
\left( 1 + \frac{ 3J'}{4J} \right)
\left( 17 + \frac{ 3J'}{4J} \right)  \right]^{1/2} }{
1+ \frac{J'}{12J} }
\end{array} \right.
\label{47}
\end{equation}

\noindent
We have assumed in (\ref{47}) that $S$ is half integer.
 From eq.(\ref{43}) we derive that $g(n_{\ell}, J'/J)$ is a monotonically
increasing function of the ratio $J'/J$, which implies that
the ladder is more disordered in the strong coupling regime
that it is for weak coupling. In fact we get,

\begin{equation}
\lim_{J'/J \rightarrow \infty} g = \left\{
\begin{array}{lc}
\sim \left\{ J'/J \right)^{1/2} \rightarrow  \infty & n_{\ell}:{\rm even}   \\
\frac{2}{S} & n_{\ell}:{\rm odd},\; S:{\rm half}\;{\rm integer} \\
\frac{\sqrt{2} }{S} & n_{\ell}:{\rm odd},\; S:{\rm integer}
\end{array}
\right.
\label{48}
\end{equation}

This equation shows that the difference between even and odd
ladders appears not only in the topological term but
also in the behaviour of $g$ as a function of the ratio
$J'/J$ in the strong coupling limit. For $n_{\ell}$ even the sigma model
enters into the strong coupling
regime $g>>1$, which is dominated by the angular momentum term
${\bf l}^2$ in the sigma model Hamiltonian (\ref{21}). Let us suppose that
we discretize the sigma model Hamiltonian at $\theta=0$ as in references
\cite{HKS}, \cite{Shan}:

\begin{equation}
H_{\sigma} = \frac{v_{\sigma}}{2}  \sum_{n}  \left[
g \; {\bf l}^2(n) -   \frac{2}{g} \;
{\bf \varphi}(n) \cdot {\bf \varphi}(n+1) + cte \right]
\label{48a}
\end{equation}

\noindent
with  ${\bf l}(n)$ satisfying the standard angular momenta algebra,
such that ${\bf l}^2$ has the spectrum $l(l+1), l=0,1, \dots$.
${\bf l}(n)$ gives the angular momenta of the
$n^{th}$ rung of the ladder. Then in the strong coupling limit the
ground state of (\ref{48a}) is obtained choosing the representation
$l =0$ for each $n$. The first excited state has $l=1$ at one site
and energy $ g v_{\sigma}$, which is the value of the gap in the limit
$g >>1$. In the case $n_{\ell} =2$ we get
from eqs (\ref{46}),

\begin{equation}
v_{\sigma } g \simeq  J',  \;\; {\rm for }\; J'/J >>1
\label{48b}
\end{equation}

The second term in (\ref{48a}) produces shifts in the ground state energy
and also delocalizes the $l=1$ excitation producing a band of states.
The gap only vanishes at $g=0$. These results agree
with the ones obtained using very different techniques namely,
numerical \cite{Dag}, \cite{Bar}
, renormalization group \cite{White}
, mean field  \cite{Gopalan},
finite size  \cite{Hatano} and bosonization
\cite{Shelton}. However in order to claim full
agreement we have also to analize what happens with the
other massive modes that we discarded in the mapping of the ladder
Hamiltonian into the sigma model Hamiltonian.
If the mass which is generated dynamically by non perturbative
effects for the field $\varphi$ is smaller than the gap
associated to the massive modes $\varphi_a$ then expect that the
map must be essentially correct, except for a finite renormalization of
the coupling constant $g$ and the spin velocity $v_{\sigma}$  
This issues  will be
considered elsewhere.

For the  odd ladders the asymptotic value of $g$ is in agreement
with (\ref{22}), in the sense that the odd ladders with spin 1/2
can be though of as single chains
with a spin 1/2 and an effective coupling constant $J_{\rm eff}$.
Indeed, for $S=1/2$ we get from (\ref{48}) that $g=4$ which is the
same value we obtain in (\ref{22}) for the single
spin 1/2 chain.

Let us consider now the weak coupling limit $J'/J << 1$.
 From (\ref{43}) and (\ref{45}) we get,

\begin{equation}
\lim_{J'/J \rightarrow 0} g =
\left\{ \begin{array}{cl}
 \sqrt{2}/\left({S} n_{\ell}\right) & n_{\ell}:
{\rm even}\;\; {\rm or}\;\; n_{\ell}:
{\rm odd}, S:{\rm integer}   \\
 & \\   {2}/\left( S \sqrt{2 n^2_{\ell} -1 } \right)
& n_{\ell}:{\rm odd},\;\; S:{\rm half} \; {\rm integer}
\end{array} \right.
\label{49}
\end{equation}

Which implies  that $g$ depends essentially on the combination
$S n_{\ell}$, as we anticipated in the discussion of eq. (\ref{29}).
The isotropic case , $J=J'$, is in fact closer to the weak coupling
values (\ref{49}) than to the strong coupling ones (\ref{48}).
Thus for $S n_{\ell} >>1$ the value of $g$ will be small and we may
use the formula $exp(- 2 \pi/g)$ to estimate the value of the energy or
mass scales of the system. This implies in particular that the mass gap
for the even ladders with $n_{\ell}$ large will decrease as
$exp(-$ cte $n_{\ell})$. This agrees at least qualitatively,
with the numerical results which give
a spin gap  $\Delta_{\rm spin} $ at the
isotropic case
equal to $0.504 J$ for $n_{\ell} = 2$ and  $\Delta_{\rm spin} \sim 0.2 J $
for $n_{\ell} =4$. Thus in the limit $n_{\ell} \rightarrow \infty$ the
gap of the even ladders should vanish exponentially. As the odd ladders
are already gapless for any number of legs one reaches in the limit 
$n_{\ell}$ the same result for both even and odd chains. However one
must
be careful in this limit since as we mentioned above the massive modes
that  we discarded in our mapping to a 1d sigma model are becoming more
important as we increase the number of legs. A more careful analysis of
this questions is needed.

\section{Final Considerations}

The application of the sigma model techniques to spin ladders
has allowed  us not only
to confirm the topological origin of the qualitatively different
behaviour of even and odd ladders, but also to get some hints
about the dependence of the physical quantities on
the values of the coupling constants $J$ and $J'$.
Much work remains to be done in this direction, but we believe
that the sigma model offers an unified, economic,
and consistent approach to
spin ladders.
The connection we have stablished in this paper  allows in principle to
apply the knowledge accumulated in the past in the study of the sigma model
to the understanding  of spin ladders.

An interesting
``recent''  result concerning the sigma models at $\theta = \pi$ is
the proof of its exact integrability \cite{Fateev},\cite{ZZ2} , which
is the
parallel of the well known
integrability of the sigma model at $\theta = 0$ \cite{ZZ1}.
The proof of integrability is done in the framework of the factorized
scattering theory. For $\theta=0$ the S-matrix is formulated for a
$O(3)$-triplet of massive particles, while for $\theta=\pi$
there are two $O(2)$-doublets of left and right moving
massless particles, whith three types of scatterings: left-left,
right-right and left-right. Using the powerful techniques of the
thermodynamic Bethe ansatz one can compute finite size effects of various
observables. In this way one can prove that the RG-flow  for $\theta=\pi$,
goes from the UV asymptotically free model
with c=2 to the IR massless $SU(2)_1 $WZW
model with c=1, as we indicated in the introduction.
The results one gets using these exact techniques
agree at the perturbative level  with the
ones obtained from the perturbation of the WZW model by the
marginal irrelevant
operator ${\bf J}_L {\bf J}_R$ \cite{FOZ}.
In this way one explains the logarithmic departure from
the scale invariant results which can then be compared with
experimental or numerical results \cite{Cardy}. These logarithmic corrections
all depend on a mass scale $\Lambda$, which is generated
dynamically in the sigma model, and which for small values of
$g$ is given essentially  by  $1/a \; exp(- 2 \pi/g)$, with $a$
the lattice spacing. In the factorized S-matrix theory the parameter
$\Lambda$ appears explicitly in the expression of the energy
and momentum of the particles. An important problem is to derive
the relation between $\Lambda$ and the microscopic parameters
of the model appearing in the Hamiltonian,
namely $n_{\ell},J$ and $J'$ \cite{Hasen}.
If the Hamiltonian happens to be integrable then one should be able
to find an exact expression for  $\Lambda$, but in general this will
not be possible and so one has to use some approximation method.
Numerical computations of
thermodynamics quantities of the spin ladders and they
comparison with the theoretical predictions may also be very useful in
stablishing this connection \cite{Troyer}.

\vspace{3cm} {\bf Acknowledgements}
I want to thank the members of the Theoretical Physics Institute
of ETH for their hospitality and specially Prof. T.M.Rice for
introducing me to the subject of spin ladders. I am also grateful
for conversations to M. Sigrist, S. Haas, B. Frischmuth,
R.Shankar, S.Sachdev, H. Tsunetsugu, M.A. Martin-Delgado, A. Berkovich,
and A. Gonzalez-Ruiz.
This work has been supported by the Swiss National Science
Foundation and by the Spanish Fund DGICYT, Ref.PR95-284.

\vspace{2cm}{\em e-mail address}:
sierra@cc.csic.es

\end{document}